\documentclass[universe,article,accept,moreauthors,pdftex]{Definitions/mdpi}

\usepackage[labelformat=simple]{subfig}

\usepackage[markup=bfit,deletedmarkup=sout]{changes}
\makeatletter
\@namedef{Changes@AuthorColor}{red}
\colorlet{Changes@Color}{red}
\makeatother
\usepackage{flushend}
\usepackage{graphicx}
\usepackage{dcolumn}
\usepackage{bm}
\usepackage{hyperref}
\usepackage{subfig}
\usepackage{float}
\usepackage{threeparttable}
\usepackage{xcolor}
\usepackage{amssymb}
\usepackage{amsmath}
\catcode`@=11
\let\savesort=\NAT@sort@cites
\newcommand\nosort[1]{\edef\NAT@cite@list{#1}}
\def\citenosort#1{\let\NAT@sort@cites=\nosort \cite{#1}%
   \let\NAT@sort@cites=\savesort}
\catcode`@=12 
\makeatletter
    
    \newcommand{\Rmnum}[1]{\expandafter\@slowromancap\romannumeral #1@}
  \makeatother

\newcommand{\hunit}{$\rm{km \ s^{-1} \ Mpc^{-1}}$}
\newcommand{\lcdm}{$\Lambda$CDM}
\usepackage{array}
\usepackage{booktabs}

\newcommand{\beq}{\begin{equation}}
\newcommand{\eeq}{\end{equation}}
\usepackage{threeparttable}
\newcommand{\bea}{\begin{eqnarray}}
\newcommand{\eea}{\end{eqnarray}}

\firstpage{1} 
\pubvolume{1}
\issuenum{1}
\articlenumber{0}
\pubyear{2021}
\copyrightyear{2021}
\externaleditor{Academic Editor: Pedro Avelino} 

\datereceived{29 January 2021} 
\dateaccepted{02 March 2021 } 
\datepublished{} 
\hreflink{https://doi.org/} 

\Title{Cosmological Constraints on the Coupling Model from Observational Hubble Parameter and Baryon Acoustic Oscillation Measurements}

\TitleCitation{Cosmological Constraints on the Coupling Model from Observational Hubble Parameter and Baryon Acoustic Oscillation Measurements}

\Author{Shulei Cao
 $^{1,2}$\orcidA{}, Tong-Jie Zhang $^{1,3,}$*, Xinya Wang $^{4}$ and Tingting Zhang $^{5}$}

\AuthorNames{Shulei Cao, Tong-Jie Zhang,  Xinya Wang and Tingting Zhang}

\AuthorCitation{Cao, S.; Zhang, T.-J.; \mbox{Wang, X.}; Zhang, T.}

\address{%
$^{1}$ \quad Department of Astronomy, Beijing Normal University, Beijing 100875, China\\
$^{2}$ \quad Department of Physics, Kansas State University, 116 Cardwell Hall, Manhattan, KS 66502, USA; shulei@phys.ksu.edu\\
$^{3}$ \quad Institute for Astronomical Science, Dezhou University, Dezhou 253023, China\\
$^{4}$ \quad Department of Industrial and Manufacturing Systems Engineering, Kansas State University, \mbox{Manhattan, KS 66506, USA;} lawrence15@ksu.edu\\
$^{5}$ \quad PLA Army Engineering University, Nanjing 210017, China; 101101964@seu.edu.cn\\
}

\corres{Correspondence: tjzhang@bnu.edu.cn}

\abstract{In the paper, we consider two models in which dark energy is coupled with either dust matter or dark matter, and discuss the conditions that allow more time for structure formation to take place at high redshifts. These models are expected to have a larger age of the universe than that of \lcdm\ [universe consists of cold dark matter (CDM) and dark energy (a cosmological constant, $\Lambda$)], so it can explain the formation of high redshift gravitationally bound systems which the \lcdm\ model cannot interpret. We use the observational Hubble parameter data (OHD) and Hubble parameter obtained from cosmic chronometers method ($H(z)$) in combination with baryon acoustic oscillation (BAO) data to constrain these models. With the best-fitting parameters, we discuss how the age, the deceleration parameter, and the energy density parameters evolve in the new universes, and compare them with that of \lcdm.}


\begin{document}
\section{Introduction}
\label{sec:introduction}
The observational developments recently
have provided cosmology with a standard model, which describes the universe by the Friedmann--Lemaitre--Robertson--Walker (FLRW) metric, with~cold pressureless matter contributing roughly 1/3 and the negative pressure dark energy (DE) contributing the remaining 2/3 of the critical energy density~\cite{Mortonson2014,Brax_2017,ARUN2017166}.

Agreeing with most observations, \lcdm\ [universe consists of cold dark matter (CDM) and dark energy (a cosmological constant, $\Lambda$)] \cite{peeb84} is
accepted by most of the scientists in cosmology. However,
recent observations provide some conflicts with \lcdm. For~example, over~400 quasars (QSOs) with $z>4$ are found, and~the seven highest redshift quasars have $z>5.7$ \cite{Sahni2005,Richards2004}. If~their centers are black holes, these black holes must have a scale of $10^9\ M_\odot$. Whether such supermassive black holes can form in
the \lcdm\ universe with an age less than $10^9$ years at
$z\sim6$ \cite{Haiman2004} remains an open question. Moreover, as~stated in~\cite{Melia2015}, the~recent discovery of SDSS 010013.02+280225.8 (hereinafter ``J0100+2802''), an~ultraluminous quasar at redshift $z=6.30$, has aggravated the problem of supermassive black-hole growth and evolution in the early Universe~\cite{Wu2015}. Not to mention the issues that the 14-Gyr age of the universe obtained by Pont~et~al.~\cite{Pont1998}, which is in tension with $13.797 \pm 0.023$ Gyr~\cite{2018arXiv180706209P}, the~3.5-Gyr-old radio galaxy 53W091 at $z = 1.55$ and 4-Gyr-old radio galaxy 53W069~\cite{Dunlop1996,Spinrad1997}. In~addition, the~existence of Pop III stars ($z\geq17$) which may have been responsible for ionizing the universe at lower redshifts is also problematic in the \lcdm\ \cite{Sahni2005}.

Motivated by the above issues and the works of Hao Wei~\cite{Wei2007o,Wei2007j,Wei2006}, here we intend to test a model of dark energy (DE) coupled with matter that ameliorates the aging problem. The~coupling between DE and dust matter has been discussed in detail~\cite{2010PhRvD..82l4006G,Aviles2011,Khoury2004}. In~the literature, the~coupling between DE and dark matter (DM) has been investigated to explain the Hubble constant ($H_0$) and $\sigma_8$ tensions~\cite{Kumar_2017,DiValentino_2017,An_2018,Yang_2018,Kumar_2019,Pan_2019,DiValentino_2020}, where $\sigma_8$ measures the amplitude of the (linear) power spectrum on the scale of 8 h$^{-1}$ Mpc, with~$h=H_0/(100\ \rm{km \ s^{-1} \ Mpc^{-1}})$ being the reduced Hubble constant. It would be interesting to also consider a model with DE and DM coupling. Therefore, for~comparison purpose, we select three coupling forms of these two models and use observational Hubble parameter data (OHD) and Hubble parameter obtained from cosmic chronometers method (denoted as $H(z)$) in combination with baryon acoustic oscillation (BAO) data to constrain the cosmological and nuisance parameters in the given models. Finally, we explore the general properties of the best-fitting models and their implications, and~determine the most favored~model.

This paper is organized as follows: In Section~\ref{sec1}, we briefly review the standard cosmology and its challenges. In~Section~\ref{sec2}, we describe the coupling models in details. The~constraints on the models are presented in
Section~\ref{sec3}. In~Section~\ref{sec4}, we discuss the properties of the best-fitting models and their merits. Finally, the~conclusions and discussions are given in Section~\ref{sec5}.

\section{The Standard Cosmology and Its~Challenges}
\label{sec1}
The current standard cosmology is mainly based on the Einstein's general relativity, and~FLRW metric, where the homogeneous and isotropic solution of Einstein's field equations, is given by
\beq
ds^{2}=-dt^{2}+a^{2}(t)[\frac{dr^{2}}{1-kr^{2}}+r^{2}d\theta^{2}+r^{2}sin^{2}\theta d\phi].
\eeq
$a(t)$ is the scale factor, $r$, $\theta$ and $\phi$ are dimensionless comoving coordinates, and~$k=0, \pm1$
 represent the curvature of the spatial Section \cite{Benoit2003}. The~homogeneous matter in the
 universe behaves as a tensor
\beq
 T^{\mu\nu}=(\rho+P)U^{\mu}U^{\nu}+Pg^{\mu\nu},
\eeq
where $\rho$ and $P$ represent the density
and the pressure of the matter, and~$U^{\mu}$ and $g_{\mu\nu}$ denote the 4-velocity of the matter and space-time metric, respectively. When U=(1, 0, 0, 0), matter is static in the reference coordinate, but~comoving with the expanding universe. From~Friedmann equations, one can get the the expression of Hubble parameter $H(z)$ in flat \lcdm, 
\beq
H^{2}=H_{0}^{2}\big(\Omega_{m_0}(1+z)^{3}+\Omega_{r_0}(1+z)^{3}+\Omega_{\Lambda}\big),
 \label{lcdm}
\eeq
where $\Omega_{m_0}$, $\Omega_{r_0}$, and~$\Omega_{\Lambda}$ are current radiation energy, current dust matter, and~constant DE density parameters, respectively, and~subscript ``0'' denotes the present value hereafter. Note that $\Omega_{r_0}$ is accurately measured, being $\Omega_{r_0}=2.47\times10^{-5} h^{-2}$ \cite{2020PTEP.2020h3C01P}, and~the current neutrino energy density parameter $\Omega_{\nu_0}$ is given by
\beq
\Omega_{\nu_0}=\frac{\sum m_{\nu}}{93.14h^2}\ ,
\label{rho_nu}
\eeq
where $\sum m_{\nu}=0.06$ eV is the total neutrino mass and the effective number of massless neutrino species $N_{\rm eff}=3.045$ \cite{2020PTEP.2020h3C01P}. Though~\lcdm\ is generally accepted by most scientists and agreed mostly with the observations, it appears that the recent observations at modest redshifts ($6\leq z\leq20$) have some surprises in store for \lcdm\ \cite{Sahni2005}.

In fact, over~400 QSOs with redshifts $z>4$ are known at present, and~the seven highest redshift quasars have $z>5.7$. If~quasars shine by virtue of an accreting black hole at their centers, then all these QSOs must host $\geq 10^{9}M_{\odot}$ black holes~\cite{Haiman2004}, and~it takes at least $7\times10^{8}$ years for such a black hole to form. We define the age of the universe at the redshifts \mbox{$z$, $t_{z}$, as}
\beq
 t_{z}=\int_{z}^{\infty}\frac{dz}{(1+z)H(z)}.
 \label{tz}
\eeq
Considering $H_0=67.4$ \hunit\ and $\Omega_{m_0}=0.315$ \cite{2018arXiv180706209P}, and~substituting Equation \eqref{lcdm} into Equation \eqref{tz}, we get the age of universe at $z\sim6$ as $9.3\times10^{8}$ years which is merely enough for the formation of $10^{9}M_{\odot}$ black~holes.

The following arguments are summarized by Melia and McClintock~\cite{Melia2015}. In~the standard context, the~universe became transparent about 0.4 Myr after the big bang, descending into the so-called Dark Ages, which ended several hundred million years later. After~that, density perturbations condensed into stars and early galaxies, producing ionizing radiation and the epoch of re-ionization (EoR) began. Standard astrophysical principles suggest that ionizing radiation was produced by Pop II and III stars. The~EoR was constrained by Zaroubi~et~al.~\cite{Zaroubi2013} at $6\lesssim z\lesssim15$.
How the universe evolved through the Dark Ages and into the EoR was studied by many detailed simulations (e.g., \cite{Bromm2009}), which show that primordial gas clouds formed in dark-matter halos with virial temperature 1000 K and mass $10^6\ M_{\odot}$ (so-called ``minihaloes''; $M_{\odot}$, solar mass). In~the standard CDM model, the~minihaloes that were the first sites for star formation are expected to be in place at redshift $z\approx 20$, when the age of the universe was just a few hundred million years~\cite{Bromm2002}. After~at least 100 Myr, 5--20 $M_{\odot}$ black-hole seeds were created, presumably following the supernova explosion of evolved Pop II (and possibly Pop III) stars. Conventionally, the~black-hole mass as a function of time can be expressed by Salpeter relation (see~\cite{Melia2015} and references therein for more details)
\beq
 M(t)=M_{\rm seed}\exp{\frac{t-t_{\rm seed}}{45\ \rm Myr}},
 \label{mt}
\eeq
where $M_{\rm seed}$ is the black-hole seed mass produced at time $t_{\rm seed}$. Note that $M(t)$, being a model-dependent quantity, may present dependence on cosmological parameters, especially on $H_0$ \cite{Nunes_2020}. Assuming $M_{\rm seed}=20\ M_{\odot}$ (upper bound), the~minimal growth time from the Salpeter relation is $t-t_{\rm seed}\sim910$ Myr for an inferred mass of approximately 10--12\,$\times 10^9\ M_{\odot}$. Since in \lcdm\ (with $H_0=67.4$ \hunit\ and $\Omega_{m_0}=0.315$ given by~\cite{2018arXiv180706209P}), $t(z=6.3)\sim873$ Myr, not only is this age of quasar J0100+2802 inconsistent with the transition from the Dark Ages to the EoR, but~also the quasar would have had to grow beyond the big bang. This is still true even if we are using the lower bounds of \mbox{$H_0=66.82$ \hunit\ and $\Omega_{m_0}=0.308$} from~\cite{2018arXiv180706209P}, which result in \mbox{$t(z=6.3)\sim891$ Myr}.

In order to improve these problems, we aim to test a new model to agree better with the observations. Indeed, a~smaller value of $H(z)$ at high redshifts can result in a larger age of the universe, so can an oscillating $H(z)$. Note that there also exist other challenges, such as $H_0$ and $\sigma_8$ tensions, but~here we mainly focus on the aging problems. We will introduce the coupling model in the following~section.

\section{Coupling between Dark Energy and Dust/Dark~Matter}
\label{sec2}
\unskip

\subsection{Dark Energy Coupled with Dust~Matter}
\label{DE+M}

The standard theory of \lcdm\ is based on the assumption that all the cosmological compositions evolve independently, while first we assume that the dust matter and DE exchange energy through interaction according to
 \beq
 \dot{\rho}_{\rm X}+3H(1+w_{\rm X})\rho_{\rm X}=-3\lambda H\rho_{m},
 \label{1}
 \eeq
 and
 \beq
 \dot{\rho}_{m}+3H\rho_{m}=3\lambda H\rho_{m},
 \label{2}
 \eeq
and the conservation equation for the radiation energy density is
 \beq
 \dot{\rho}_{r}+4H\rho_{r}=0,
 \label{3}
 \eeq
 which preserves the total energy conservation equation
 \mbox{$\dot{\rho}_{\rm tot}+3H(\rho_{\rm tot}+P_{\rm tot})=0$ \cite{Wei2007o,Wei2007j,Wei2006},} where $\lambda$, ${\rho}_{m}$, ${\rho}_{\rm X}$, ${\rho}_{r}$, and~$\dot{\rho}$ denote the coupling coefficient, the~matter density, the~DE density, the~radiation energy density, and~their time derivatives. Some researchers use this model to discuss the Chameleon DE and its test of the solar system~\cite{2010PhRvD..82l4006G,Aviles2011,Khoury2004}. Here, $\lambda$ determines the coupling form between the matter and DE, and~the DE equation of state (EoS) $w_{\rm X}=P_{\rm X}/\rho_{\rm X}$ is constant. Here we only consider the spatially-flat FLRW universe. With~$3\lambda H\rho_{m}$ as the interaction term, here we assume that the coupling coefficient takes the form  $\lambda=A_1^3\cos\big[\frac{\ln(1+z)}{A_2}-A_1\pi\big]$ \cite{Wei2007o}, in~which $A_1$ and $A_2$ are undetermined constant coefficients. Substituting the expression of $\lambda$ into Equation~(\ref{2}), we can get the \mbox{expression of $\rho_{m}$,}
 \beq
 \rho_{m}(z)=\rho_{m_0}(1+z)^3\exp\Bigg\{-3A_1^3A_2\Bigg[\sin(A_1\pi)+\sin\bigg[\frac{\ln(1+z)}{A_2}-A_1\pi\bigg]\Bigg]\Bigg\},
 \label{rhom2}
 \eeq
 where $\rho_{m_0}\equiv\rho_{m}(z=0)$.
 Substituting Equation~(\ref{rhom2}) into Equation~(\ref{1}), we can obtain the \mbox{expression of $\rho_{\rm X}$,}
\beq
\rho_{\rm X}(z)=(1+z)^{3(1+w_{\rm X})}\big[\rho_{\rm X_0}+\rho_{m_0}I(z)\big],
\label{rhox}
\eeq
where
\beq
\begin{aligned}
I(z)=&\int^{z}_{0}\frac{3A_1^3}{(1+x)^{3w_{\rm X}+1}}\cos\bigg[\frac{\ln(1+x)}{A_2}-A_1\pi\bigg]\\
&\exp\Bigg\{-3A_1^3A_2\Bigg[\sin\bigg[\frac{\ln(1+x)}{A_2}-A_1\pi\bigg]+\sin(\frac{\pi}{A_1})\Bigg]\Bigg\}dx.
\end{aligned}
\label{Iz}
\eeq
Similarly, from Equation \eqref{3}, we can get
\beq
\rho_{r}(z)=\rho_{r_0}(1+z)^4.
\label{rr}
\eeq

Therefore in this model, we~have

 \beq
 \frac{\rho_{m}+\rho_{\rm X}+\rho_{r}}{\rho_{\rm crit}}=\Omega_{m}(z)+\Omega_{\rm X}(z)+\Omega_{r}(z)=1,
 \label{midu}
 \eeq
in which $\Omega_{m}=\rho_{m}/\rho_{\rm crit}$, $\Omega_{\rm X}=\rho_{\rm X}/\rho_{\rm crit}$, and~$\Omega_{r}=\rho_{r}/\rho_{\rm crit}$ are the dust matter density, DE density, and~radiation energy density parameters, respectively, with~$\rho_{\rm crit}\equiv3H^2/8\pi G$ being the critical density. Furthermore, $\Omega_{m_0}+\Omega_{\rm X_0}+\Omega_{r_0}=1$ is satisfied. Then, the~Friedmann equation becomes
\beq
\begin{aligned}
\frac{H^{2}}{H_{0}^{2}}=&\Omega_{m_0}(1+z)^3\exp\Bigg\{-3A_1^3A_2\Bigg[\sin(A_1\pi)+\sin\bigg[\frac{\ln(1+z)}{A_2}-A_1\pi\bigg]\Bigg]\Bigg\}\\
&+(1+z)^{3(1+w_{\rm X})}\bigg\{1-\Omega_{\rm m_0}-\Omega_{r_0}+\Omega_{m_0}I(z)\bigg\}+\Omega_{r_0}(1+z)^4.
\end{aligned}
\label{fe}
\eeq

This equation is the $H(z)$ expression of this coupling model (denoted as ``Model 1'' hereinafter).

We could also simply put $\lambda=$ const. In~this case (denoted as ``Model 2'' hereinafter), the~Hubble parameter becomes
 \begingroup\makeatletter\def\f@size{9.5}\check@mathfonts
\def\maketag@@@#1{\hbox{\m@th\normalsize\normalfont#1}}%
\beq
H(z)=H_0\sqrt{\frac{w_{\rm X}\Omega_{m_0}}{\lambda+w_{\rm X}}(1+z)^{3(1-\lambda)}+\Big(1-\Omega_{r_0}-\frac{w_{\rm X}\Omega_{m_0}}{\lambda+w_{\rm X}}\Big)(1+z)^{3(1+w_{\rm X})}+\Omega_{r_0}(1+z)^4}.
\label{hz1}
\eeq
\endgroup
\subsection{{Dark Energy Coupled with Dark~Matter}}
\label{DE+DM}

{If we consider the coupling between DE and (cold) DM (see~\cite{Bolotin_2015} for review), the~interaction equations read
\beq
\dot{\rho}_{\rm X}+3H(1+w_{\rm X})\rho_{\rm X}=-3\lambda H\rho_{X},
\label{r1}
\eeq
and
\beq
\dot{\rho}_{c}+3H\rho_{c}=3\lambda H\rho_{X},
\label{r2}
\eeq
where $\rho_{c}$ is the energy density for cold DM. Here we also consider $\lambda=$ const. Consequently, we can obtain the expressions of $\rho_{\rm X}$
\beq
\rho_{\rm X}(z)=\rho_{\rm X_0}(1+z)^{3(1+w_{\rm X}+\lambda)},
\label{rx2}
\eeq
and $\rho_{c}$
\beq
\rho_{c}(z)=(1+z)^3\bigg\{\rho_{c_0}+\frac{\lambda}{w_{\rm X}+\lambda}\rho_{\rm X_0}\Big[(1+z)^{3(w_{\rm X}+\lambda)}-1\Big]\bigg\},
\label{rc2}
\eeq
where $\rho_{\rm X_0}\equiv\rho_{\rm X}(z=0)$ and $\rho_{c_0}\equiv\rho_{c}(z=0)$.
The conservation equation for the energy density of the barynoic matter is
\beq
\dot{\rho}_{b}+3H\rho_{b}=0,
\label{r3}
\eeq
which leads to
\beq
\rho_{b}(z)=\rho_{b_0}(1+z)^3.
\label{rb}
\eeq

In this case {(denoted as ``Model 3'' hereinafter), the~dust matter density becomes \mbox{$\rho_{m}=\rho_{b}+\rho_{c}$}.} Following the same definitions of the cosmological density parameters as above, $\Omega_{b}$, $\Omega_{c}$, and~$\Omega_{\rm X}$ are the baryon, DM, and~DE density parameters, respectively, and~$\Omega_{b}+\Omega_{c}+\Omega_{\rm X}+\Omega_{r}=1$. The~Friedman equation then becomes
 \begingroup\makeatletter\def\f@size{9}\check@mathfonts
\def\maketag@@@#1{\hbox{\m@th\normalsize\normalfont#1}}%
\beq
H(z)=H_0\sqrt{(1+z)^3\bigg\{\Omega_{b_0}+\Omega_{c_0}+\frac{\Omega_{\rm X_0}}{w_{\rm X}+\lambda}\Big[(w_{\rm X}+2\lambda)(1+z)^{3(w_{\rm X}+\lambda)}-\lambda\Big]+\Omega_{r_0}(1+z)\bigg\}},
\label{hz3}
\eeq
\endgroup
where $\Omega_{m_0}=\Omega_{b_0}+\Omega_{c_0}$ and $\Omega_{m_0}+\Omega_{\rm X_0}+\Omega_{r_0}=1$.}

In the following section, we will constrain these models with OHD and $H(z)$ + BAO~combination.

\section{Constraints on the Coupling Model with Observational~Data}
\label{sec3}
 The 40 OHD we used here are based on cosmic chronometers (see~\cite{Ryan_1,Ryan_2,Caoetal_2020,Cao_2020,Cao2021} for the usage of these 31 $H(z)$ data) and radial BAO size methods, as~shown in Table~\ref{tab1}, where we could use them together is because they are statistically-independent (systematic uncertainty of individual data point is accounted for, and~in fact, based on the reduced $\chi^2$ value listed in Table~\ref{table2}, the~errors are overestimated). Systematic errors that affect $H(z)$ measurements from cosmic chronometers were studied~\cite{moresco_et_al_2012, moresco_et_al_2016}, and~were recently re-examined in~\cite{moresco_et_al_2018,moresco_et_al_2020}. Note that the statistically-dependent (correlated) data~\cite{Alam2017} has been removed (which is included in BAO measurements) compared with what used in~\cite{Cao2018} (see also~\cite{Cao2018a,Cao2018b} for more details regarding OHD).

The 11 BAO data used here are the same as what used in Cao~et~al. (see~\cite{Cao2021} for more details). The~systematic uncertainties of these BAO data are examined and found to be either negligible compared with the statistical uncertainties or included in the corresponding covariance~matrix.

\begin{specialtable}[H]
\centering
\begin{threeparttable}
\caption{The current available OHD~dataset.}\label{tab1}
\setlength{\tabcolsep}{6.1mm}{
\begin{tabular}{lccc}
\toprule
{\boldmath{$z$}}   & \textbf{\boldmath{$H(z)$} \tnote{a}} & \textbf{Method \tnote{b}} & \textbf{Ref.}\\
\midrule
$0.0708$   &  $69.0\pm19.68$      &  I    &  Zhang~et~al. (2014)-\cite{Zhang2014}   \\
      $0.09$       &  $69.0\pm12.0$        &  I    &  Jimenez~et~al. (2003)-\cite{Jimenez2003}   \\
      $0.12$       &  $68.6\pm26.2$        &  I    &  Zhang~et~al. (2014)-\cite{Zhang2014}   \\
      $0.17$       &  $83.0\pm8.0$          &  I    &  Simon~et~al. (2005)-\cite{Simon2005}     \\
      $0.179$     &  $75.0\pm4.0$          &  I    &  Moresco~et~al. (2012)-\cite{Moresco2012}     \\
      $0.199$     &  $75.0\pm5.0$          &  I    &  Moresco~et~al. (2012)-\cite{Moresco2012}     \\
      $0.2$         &  $72.9\pm29.6$        &  I    &  Zhang~et~al. (2014)-\cite{Zhang2014}   \\
      $0.240$     &  $79.69\pm2.65$      &  II   &  Gazta$\tilde{\rm{n}}$aga~et~al. (2009)-\cite{Gaztanaga2009}   \\
      $0.27$       &  $77.0\pm14.0$        &  I    &    Simon~et~al. (2005)-\cite{Simon2005}   \\
      $0.28$       &  $88.8\pm36.6$        &  I    &  Zhang~et~al. (2014)-\cite{Zhang2014}   \\
      $0.35$       &  $84.4\pm7.0$          &  II   &   Xu~et~al. (2013)-\cite{Xu2013}  \\
      $0.352$     &  $83.0\pm14.0$        &  I    &  Moresco~et~al. (2012)-\cite{Moresco2012}   \\
      $0.3802$     &  $83.0\pm13.5$        &  I    &  Moresco~et~al. (2016)-\cite{moresco_et_al_2016}   \\
      $0.4$         &  $95\pm17.0$           &  I    &  Simon~et~al. (2005)-\cite{Simon2005}     \\
      $0.4004$     &  $77.0\pm10.2$        &  I    &  Moresco~et~al. (2016)-\cite{moresco_et_al_2016}   \\
      $0.4247$     &  $87.1\pm11.2$        &  I    &  Moresco~et~al. (2016)-\cite{moresco_et_al_2016}   \\
      $0.43$     &  $86.45\pm3.68$        &  II   &  Gazta$\tilde{\rm{n}}$aga~et~al. (2009)-\cite{Gaztanaga2009}   \\
      $0.44$       & $82.6\pm7.8$           &  II   &  Blake~et~al. (2012)-\cite{Blake2012}  \\
      $0.4497$     &  $92.8\pm12.9$        &  I    &  Moresco~et~al. (2016)-\cite{moresco_et_al_2016}   \\
      $0.47$      &   $89\pm50$           &   I    &  Ratsimbazafy~et~al. (2017)-\cite{Ratsimbazafy2017}    \\
      $0.4783$     &  $80.9\pm9.0$        &  I    &  Moresco~et~al. (2016)-\cite{moresco_et_al_2016}   \\
      $0.48$       &  $97.0\pm62.0$        &  I    &  Stern~et~al. (2010)-\cite{Stern2010}     \\
      $0.57$       &  $92.4\pm4.5$          &  II   &  Samushia~et~al. (2013)-\cite{Samushia2013}   \\
      $0.593$     &  $104.0\pm13.0$      &  I    &  Moresco~et~al. (2012)-\cite{Moresco2012}   \\
      $0.6$         &  $87.9\pm6.1$          &  II   &  Blake~et~al. (2012)-\cite{Blake2012}   \\
      $0.68$       &  $92.0\pm8.0$          &  I    &  Moresco~et~al. (2012)-\cite{Moresco2012}   \\
      $0.73$       &  $97.3\pm7.0$          &  II   &  Blake~et~al. (2012)-\cite{Blake2012}  \\
      $0.781$     &  $105.0\pm12.0$      &  I    &  Moresco~et~al. (2012)-\cite{Moresco2012}   \\
      $0.875$     &  $125.0\pm17.0$      &  I    &  Moresco~et~al. (2012)-\cite{Moresco2012}   \\
      $0.88$       &  $90.0\pm40.0$        &  I    &  Stern~et~al. (2010)-\cite{Stern2010}     \\
      $0.9$         &  $117.0\pm23.0$      &  I    &  Simon~et~al. (2005)-\cite{Simon2005}  \\
      $1.037$     &  $154.0\pm20.0$      &  I    &  Moresco~et~al. (2012)-\cite{Moresco2012}   \\
      $1.3$         &  $168.0\pm17.0$      &  I    &  Simon~et~al. (2005)-\cite{Simon2005}     \\
      $1.363$     &  $160.0\pm33.6$      &  I    &  Moresco (2015)-\cite{Moresco2015}  \\
      $1.43$       &  $177.0\pm18.0$      &  I    &  Simon~et~al. (2005)-\cite{Simon2005}     \\
      $1.53$       &  $140.0\pm14.0$      &  I    &  Simon~et~al. (2005)-\cite{Simon2005}     \\
      $1.75$       &  $202.0\pm40.0$      &  I    &  Simon~et~al. (2005)-\cite{Simon2005}     \\
      $1.965$     &  $186.5\pm50.4$      &  I    &   Moresco (2015)-\cite{Moresco2015}  \\
      $2.34$       &  $222.0\pm7.0$        &  II   &  Delubac~et~al. (2015)-\cite{Delubac2015}   \\
      $2.36$       &  $226.0\pm8.0$       &   II   &  Font-Ribera~et~al. (2014)-\cite{Font-Ribera2014}    \\
\bottomrule
\end{tabular}}
\pbox{\columnwidth}{\footnotesize \tnote{a} \hunit. \tnote{b} Methods I and II represent the cosmic chronometers and the radial BAO size methods, respectively.}
\end{threeparttable}
\end{specialtable}

\begin{figure}[H]
\widefigure
\centering
  \subfloat[]{%
    \includegraphics[width=3.5in,height=3.0in]{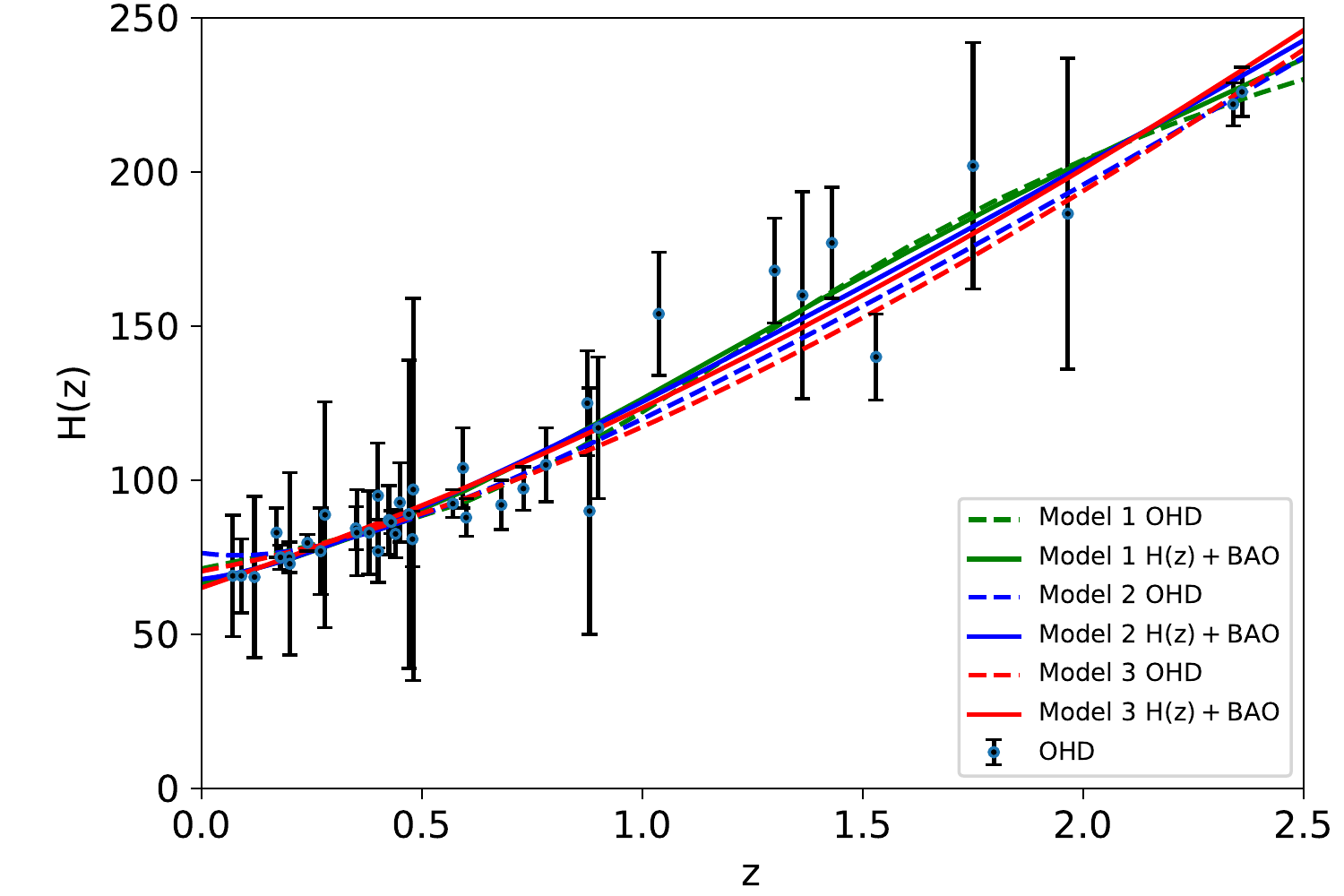}}
  \subfloat[]{%
    \includegraphics[width=3.5in,height=3.0in]{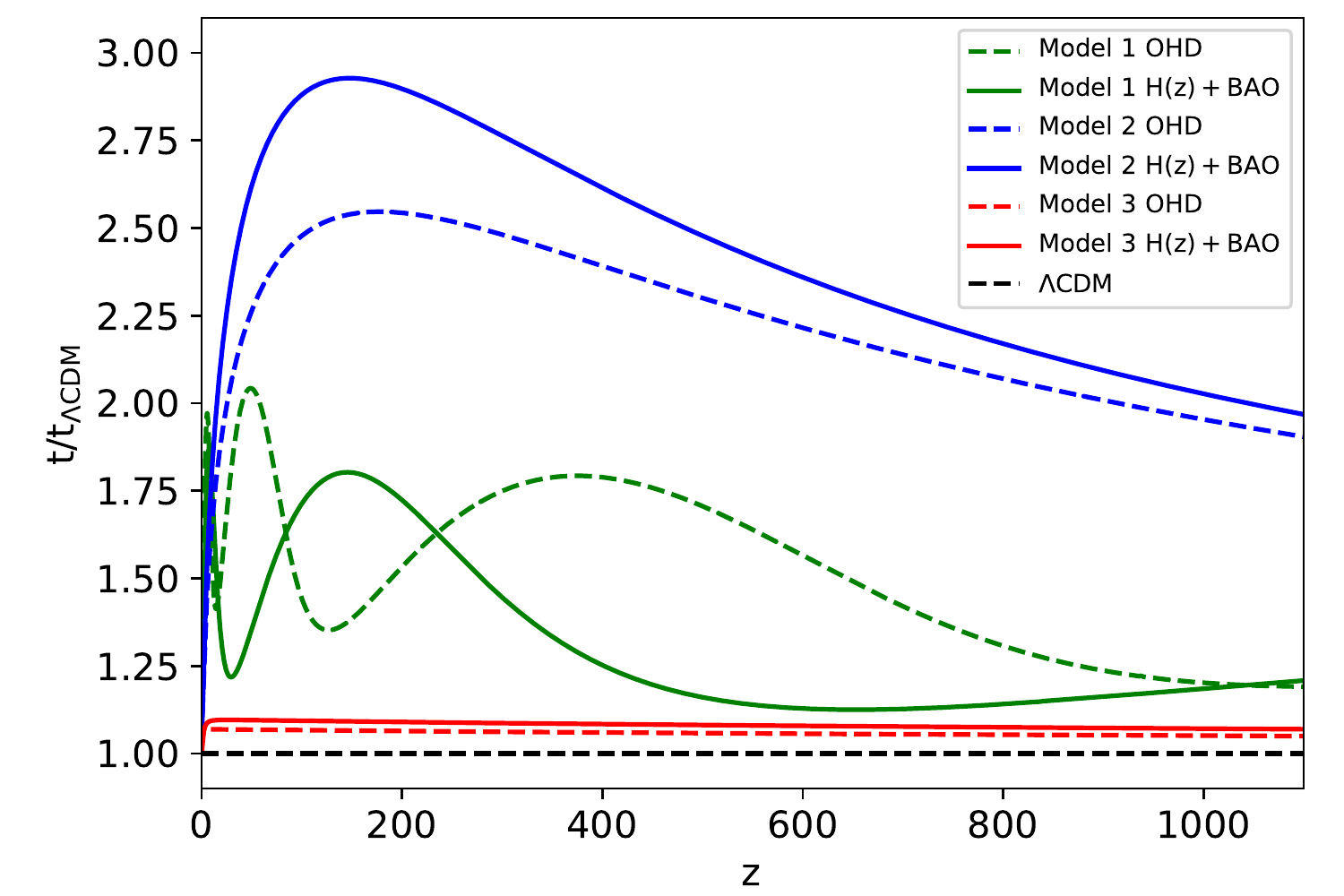}}\\
\caption{Hubble parameter $H(z)$ as a function of redshift $z$ for the coupling models with best-fitting parameters listed in Table~\ref{table2}, along with OHD (left panel), and~the age ratio $t/t_{\Lambda \rm CDM}$ as a function of redshift $z$ (right panel), where the current values are 1.11 (1.08), 1.04 (1.04), and~1.02 (1.01) for Model 1, 2, and~3 constrained from OHD ($H(z)$ + BAO), respectively.} \label{fig1}
\end{figure}

\begin{figure}[H]
\widefigure
    \includegraphics[width=6.5in,height=6.5in]{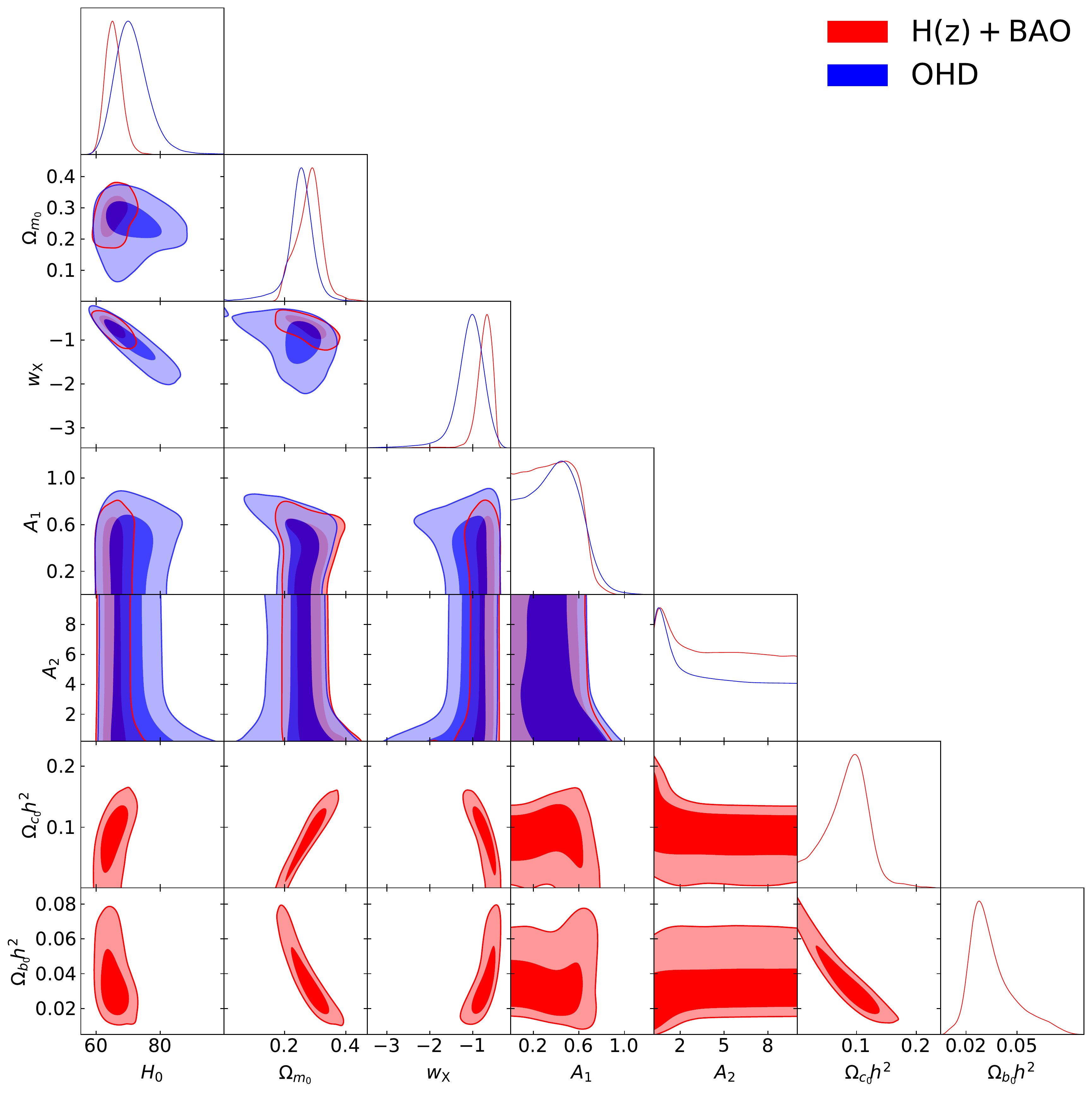}
    \caption{1$\sigma$ and 2$\sigma$ contours for Model 1 with different data combinations with best-fitting results shown in Tables~\ref{table2} and \ref{table3}.}
    \label{fig2}
\end{figure}

\begin{figure}[H]
\widefigure
    \centering
    \includegraphics[width=6.5in,height=6.5in]{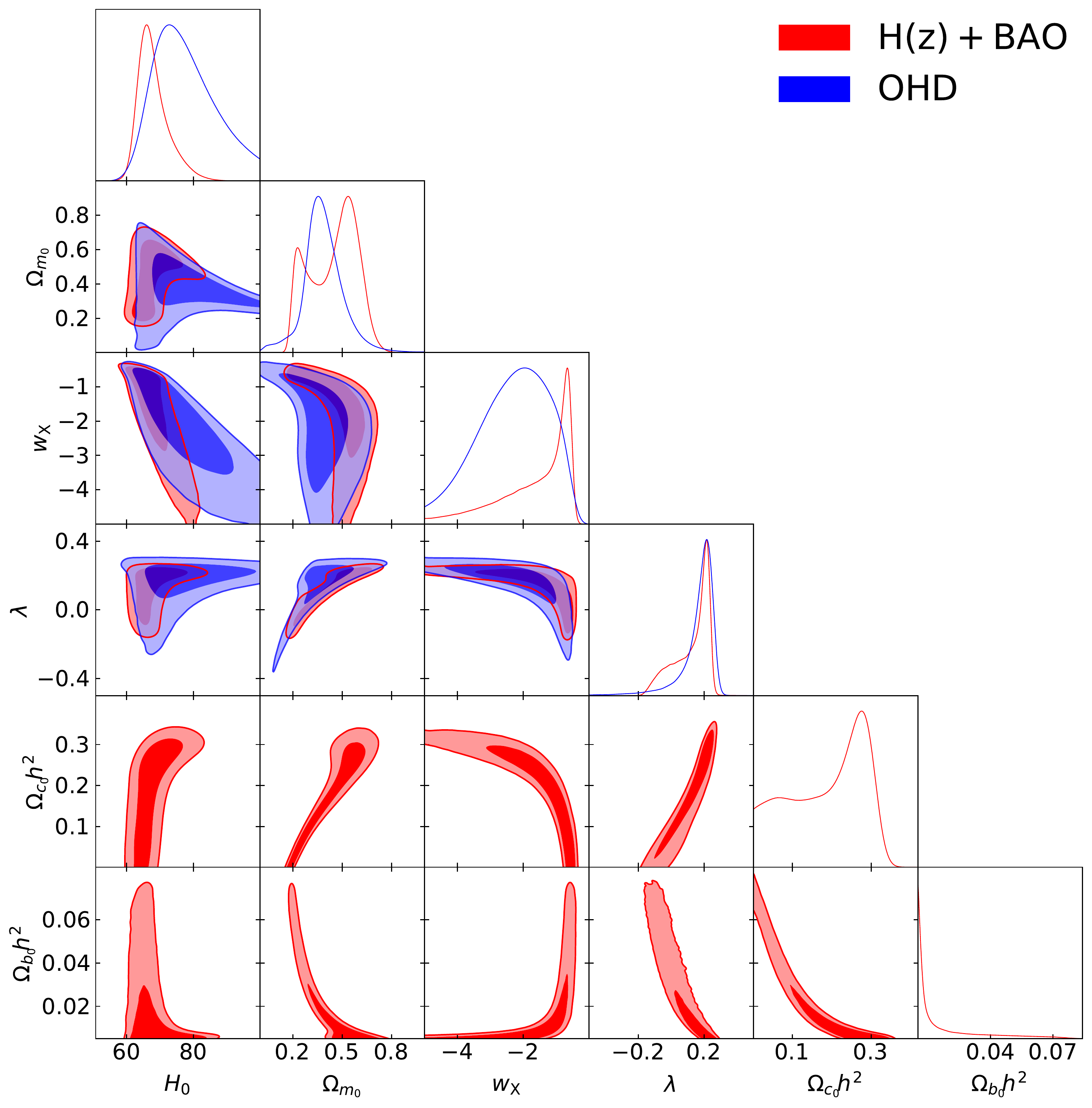}
    \caption{Same as Figure~\ref{fig2}, but~for Model~2.}
    \label{fig3}
\end{figure}

\begin{figure}[H]
\widefigure
    \centering
    \includegraphics[width=6.5in,height=6.5in]{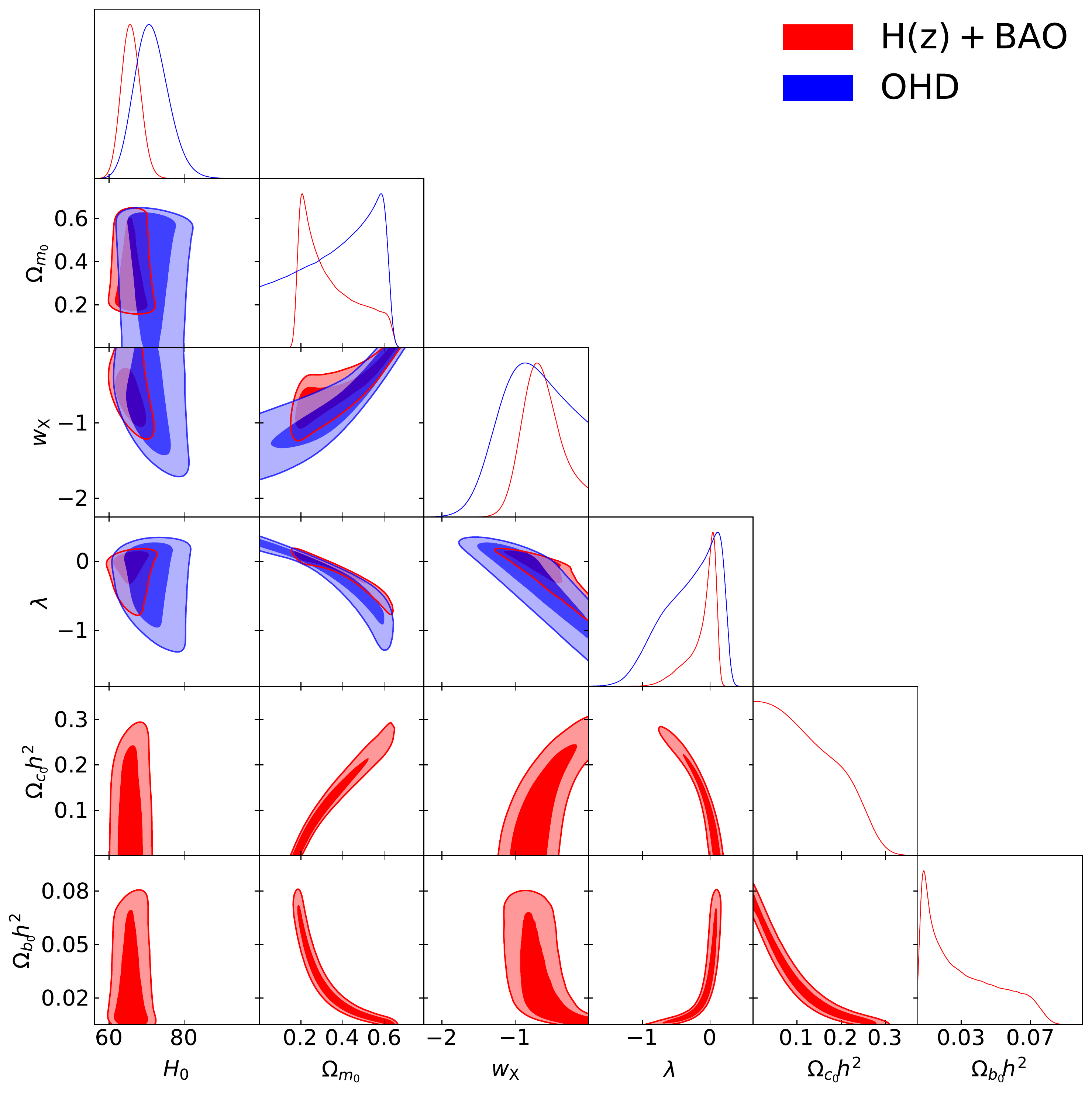}
    \caption{Same as Figure~\ref{fig2}, but~for Model~3.}
    \label{fig4}
\end{figure}

In consideration of the non-Gaussian posterior in parameter space, we perform a full Markov chain Monte Carlo (MCMC) Metropolis-Hastings sampling with emcee~\cite{2013PASP..125..306F} and assume flat parameter priors to constrain these parameters, for~OHD (or $H(z)$ data)
 \beq
  \chi^2=\sum_i\frac{[H_{\rm th}(z_i|\textit{\textbf{p}})-H_{\rm obs}(z_i)]^2}{\sigma^2(z_i)},
 \label{xh}
 \eeq
 in which $H_{\rm th}(z_i|\textit{\textbf{p}})$, $H_{\rm obs}(z_i)$, and~$\sigma(z_i)$ are the theoretical Hubble parameter at redshift $z_i$, the~OHD (or $H(z)$ data), and~the uncertainty of each $H_{\rm obs}(z_i)$, respectively. For~BAO, we use the same procedure given in Cao~et~al.~\cite{Caoetal_2020}. We show in Tables~\ref{table2} and \ref{table3} the unmarginalized and marginalized best-fitting results (the marigninalized ones are the posterior means) with 1$\sigma$ confidence regions, and~with the best-fitting parameters (details described below), we can substitute Equation~(\ref{fe}) into Equation~(\ref{tz}) to obtain the age of the universe. Furthermore, panal (a) of Figure~\ref{fig1} shows theoretical curves of $H(z)$ for the best-fitting models along with the OHD. The~1$\sigma$ and 2$\sigma$ contours are shown in Figures~\ref{fig2}--\ref{fig4}, which are analyzed by using the Python package GetDist~\cite{Lewis_2019}. {Note that in Model 1, $\textit{\textbf{p}}=\{H_0, \Omega_{m_0}, w_{\rm X}, A_1, A_2\}$ for OHD and $\textit{\textbf{p}}=\{H_0, \Omega_{c_0}\!h^2, \Omega_{b_0}\!h^2, w_{\rm X}, A_1, A_2\}$ for $H(z)$ + BAO combination; in Models 2 and 3, $\textit{\textbf{p}}=\{H_0, \Omega_{m_0}, w_{\rm X}, \lambda\}$ for OHD and $\textit{\textbf{p}}=\{H_0, \Omega_{c_0}\!h^2, \Omega_{b_0}\!h^2, w_{\rm X}, \lambda$\} for $H(z)$ + BAO combination; the priors on $\textit{\textbf{p}}$ are non-zero and flat over the ranges $0.005 \leq \Omega_{\rm b_0}\!h^2 \leq 0.1$, $0.001 \leq \Omega_{\rm c_0}\!h^2 \leq 0.99$, $0.2 \leq h \leq 1.0$, $-5 \leq w_{\rm X} \leq 0$, $0 \leq A_1 \leq 2$, $0.2 \leq A_2 \leq 10$, and~$-5 \leq \lambda \leq 5$. We also include the Akaike Information Criterion ($AIC$) and the Bayesian Information Criterion ($BIC$) to compare the goodness of fit of these models with different numbers of parameters, where
\beq
\label{AIC}
    AIC=\chi^2_{\rm min}+2n,
\eeq
and
\beq
\label{BIC}
    BIC=\chi^2_{\rm min}+n\ln N,
\eeq
where $n$ and $N$ are the numbers of free parameters of the given model and of data points. For~OHD, $N=40$, and~for $H(z)$ + BAO combination, $N=42$. The~degrees of freedom is $\nu=N-n$, which is used to determine the reduced $\chi^2$, i.e.,~$\chi^2_{\rm min}/\nu$ listed in Table~\ref{table2}.} We will discuss the details of these best-fitting models in detail in the following~section.

\begin{figure}[H]
\widefigure
\centering
  \subfloat[]{%
    \includegraphics[width=3.5in,height=3.0in]{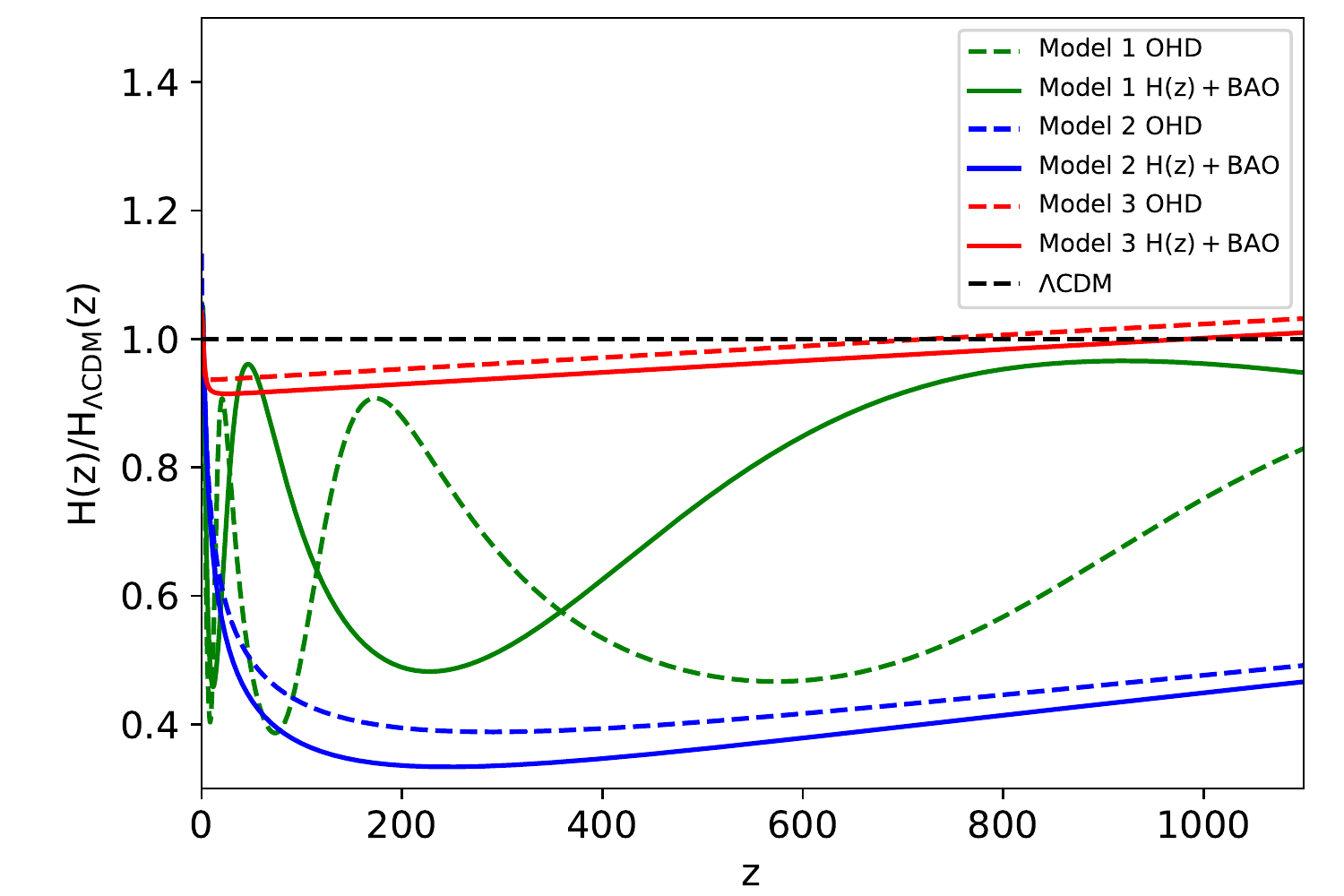}}
  \subfloat[]{%
    \includegraphics[width=3.5in,height=3.0in]{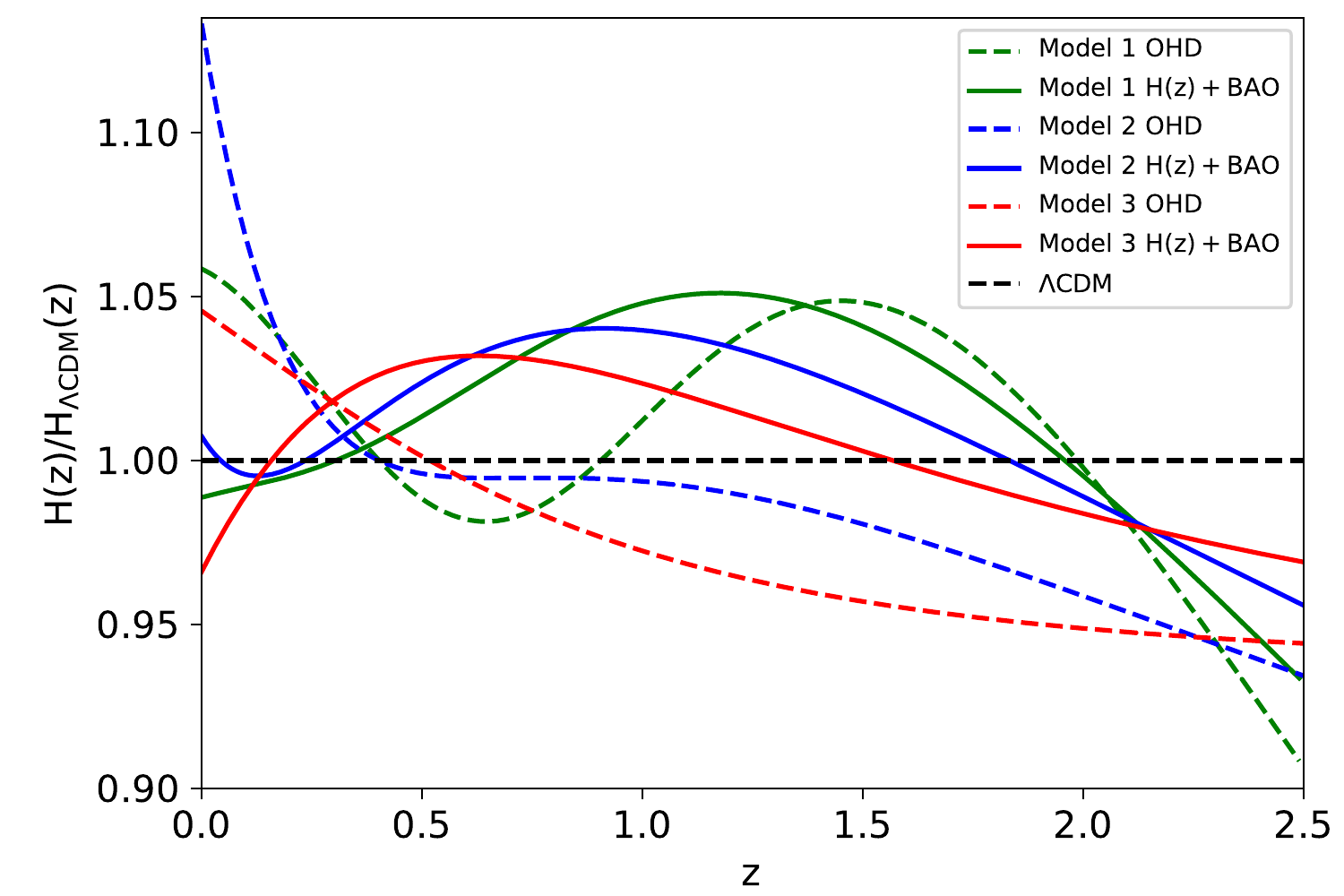}}
  \caption{The ratio between Hubble parameter $H(z)$ of coupling models and that of \lcdm\ against redshift $z$ and the right panel shows that of redshift range $0\leq z\leq 2.5$.}
\label{fig5}
\end{figure}

\begin{figure}[H]
\widefigure
  \subfloat[]{%
    \includegraphics[width=3.5in,height=3.0in]{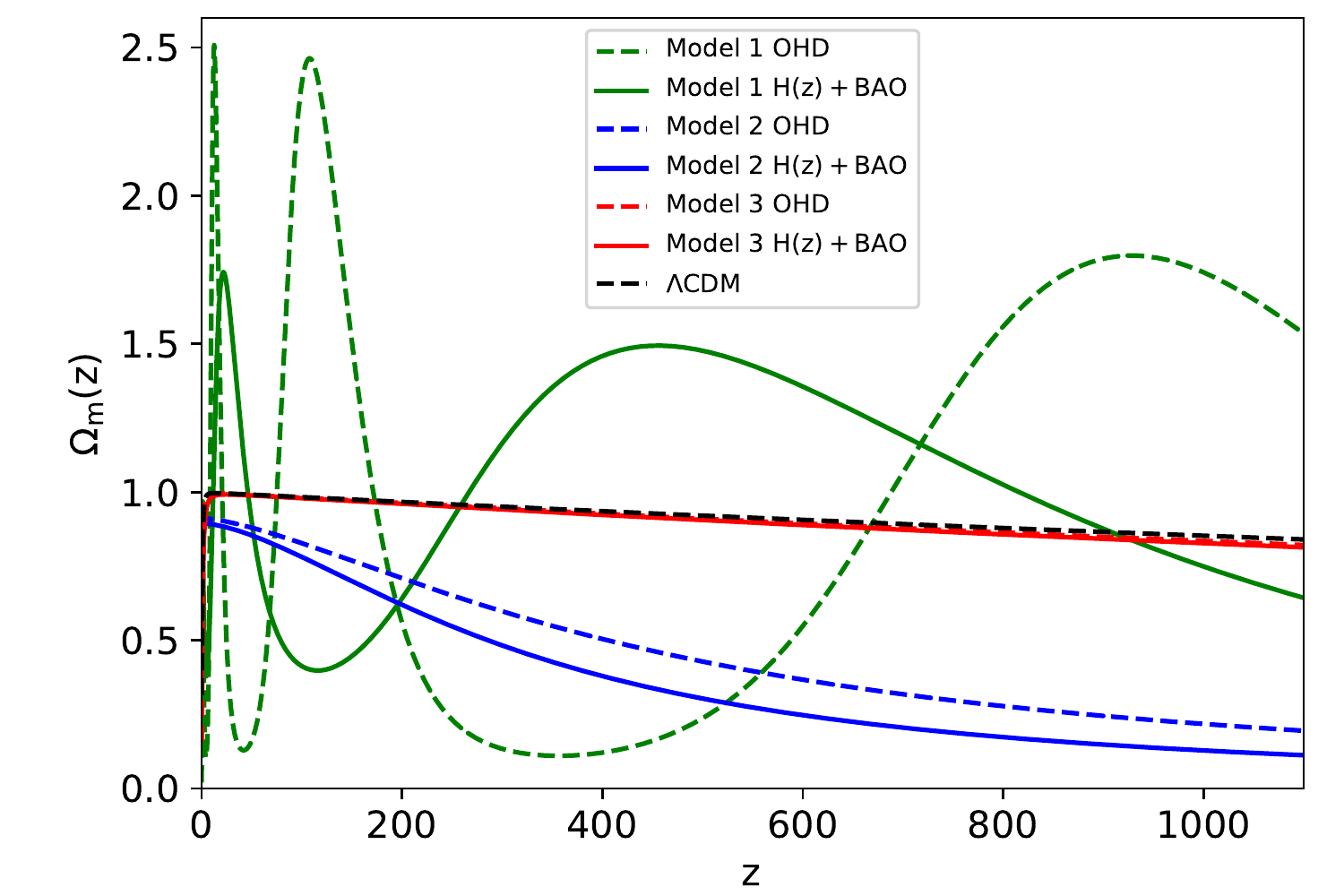}}
  \subfloat[]{%
    \includegraphics[width=3.5in,height=3.0in]{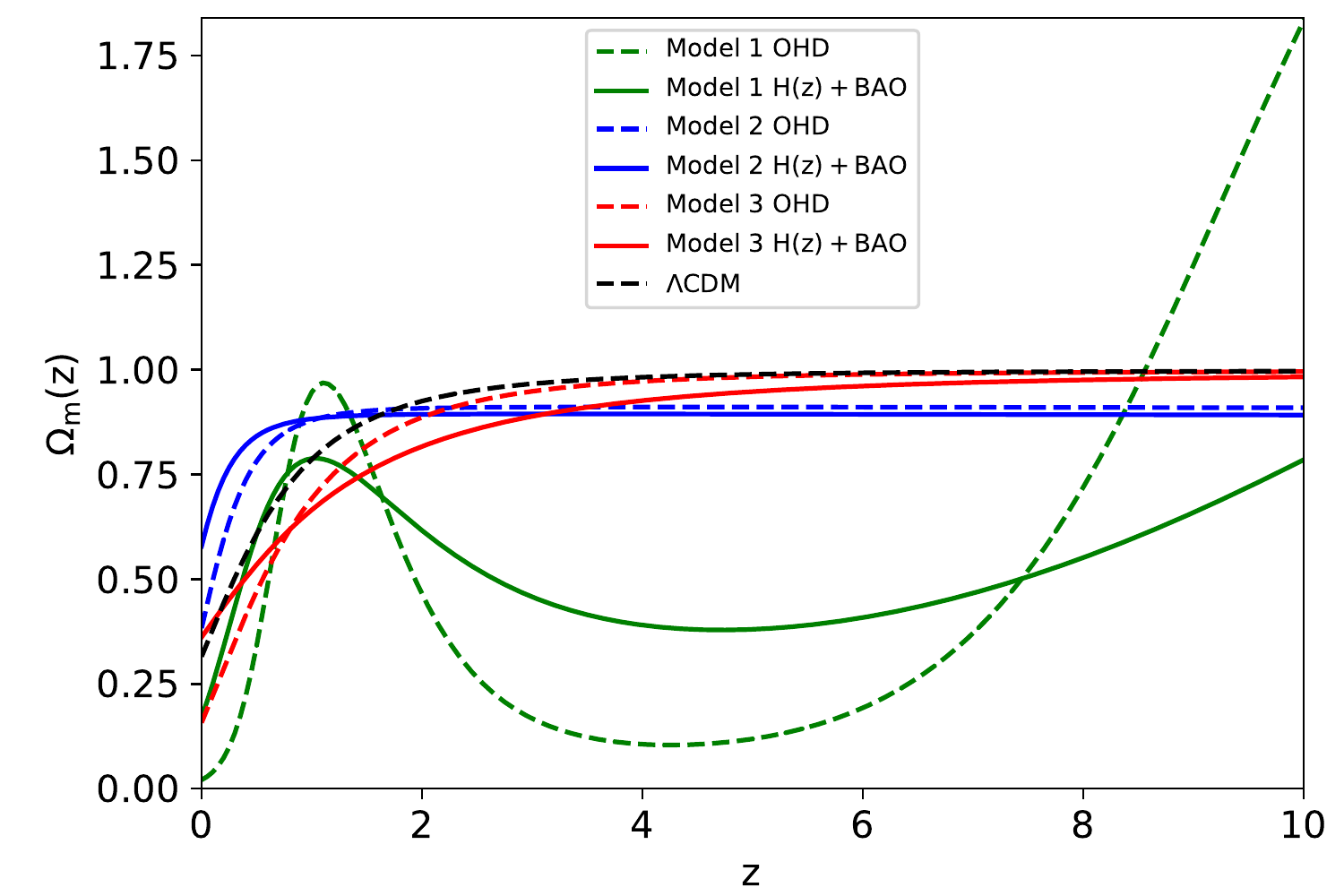}}
  \caption{The cosmological matter density parameter $\Omega_{m}$ as a function of redshift $z$ and the right panel shows that of redshift range $0\leq z\leq 10$.}
\label{fig6}
\end{figure}

\begin{figure}[H]
\widefigure
\centering
  \subfloat[]{%
    \includegraphics[width=3.5in,height=3.0in]{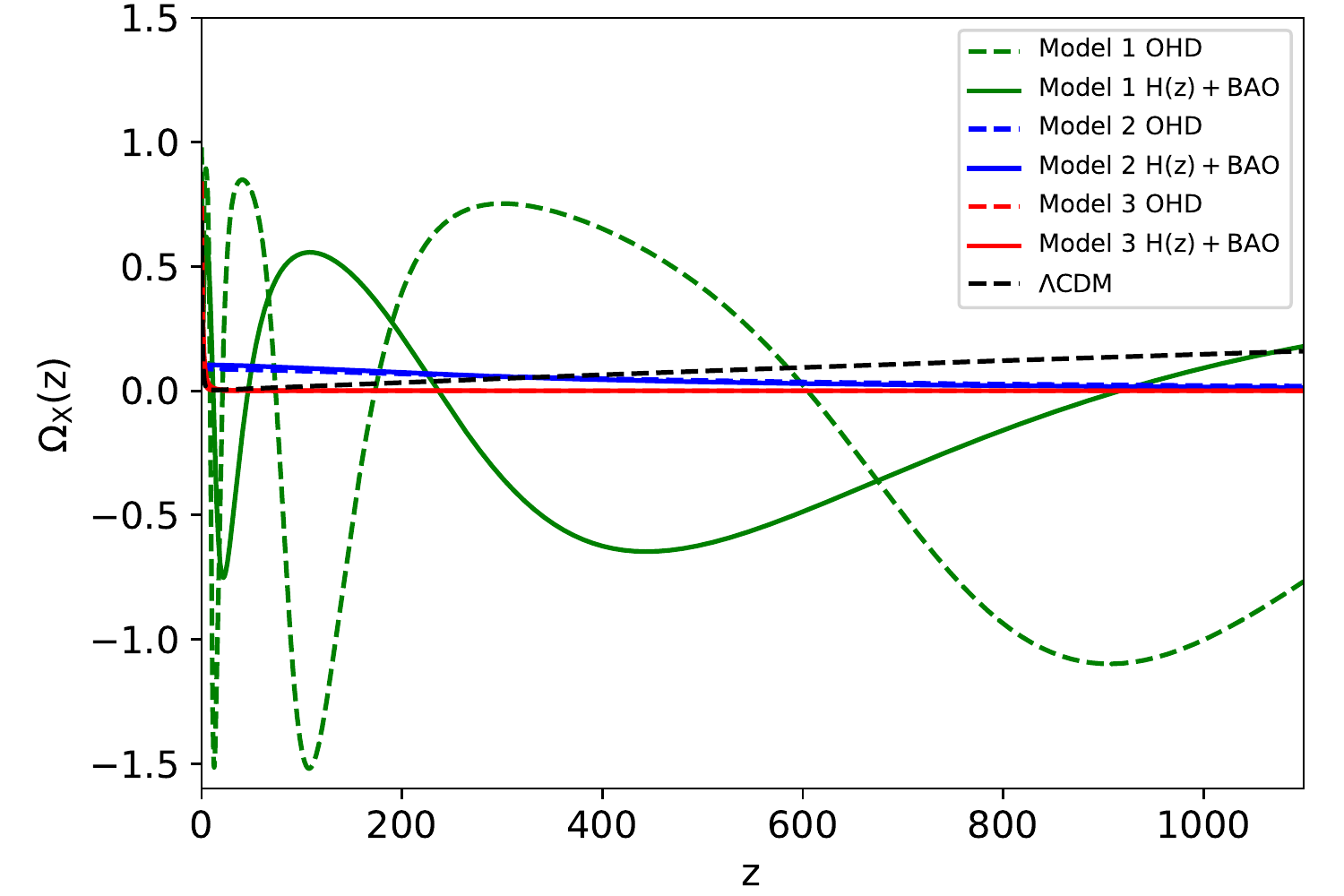}}
  \subfloat[]{%
    \includegraphics[width=3.5in,height=3.0in]{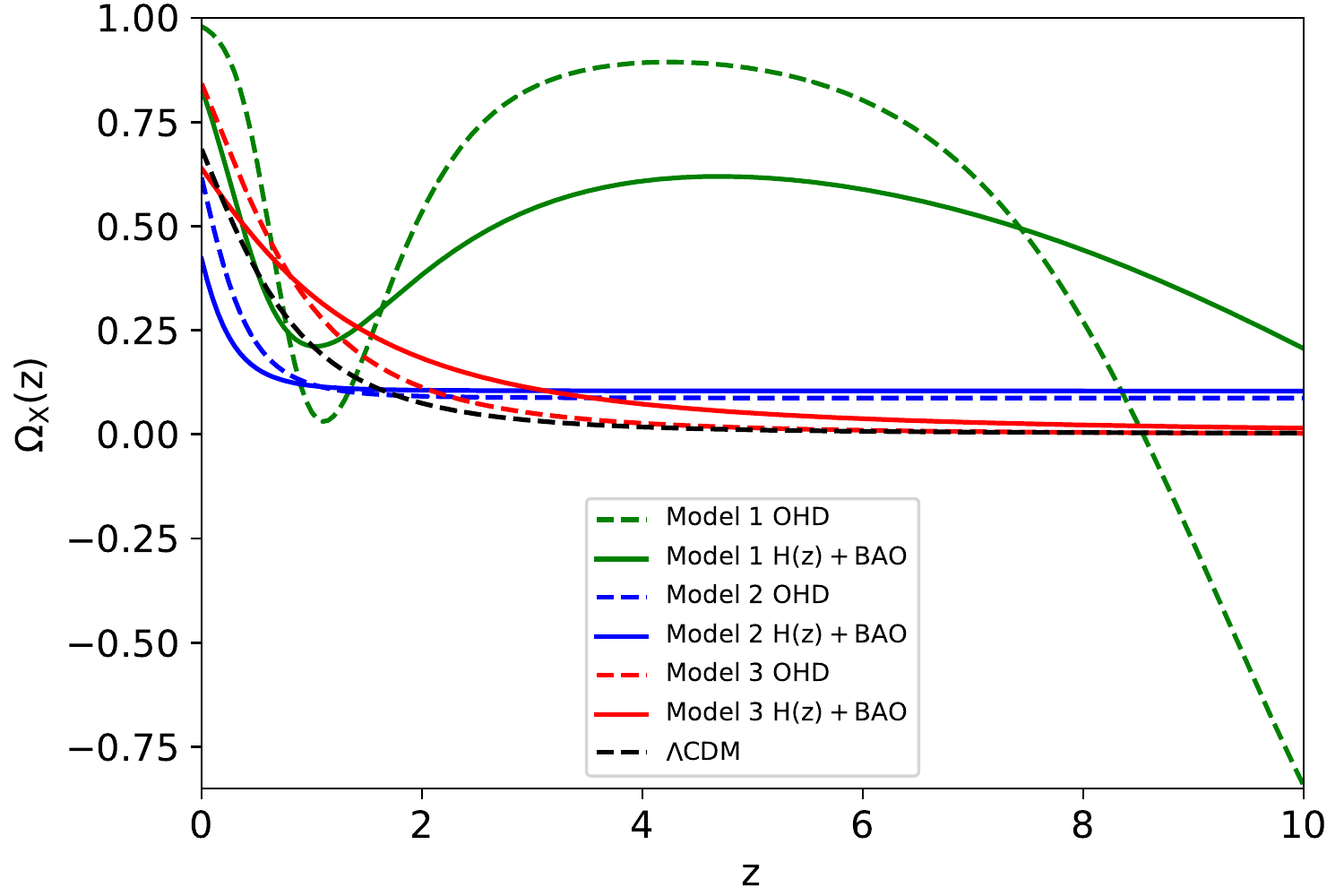}}
  \caption{The cosmological dark energy density parameter $\Omega_{\rm X}$ as a function of redshift $z$ and the right panel shows that of redshift range $0\leq z\leq 10$.}
\label{fig7}
\end{figure}

\begin{figure}[H]
\widefigure
\centering
  \subfloat[]{%
    \includegraphics[width=3.5in,height=3.0in]{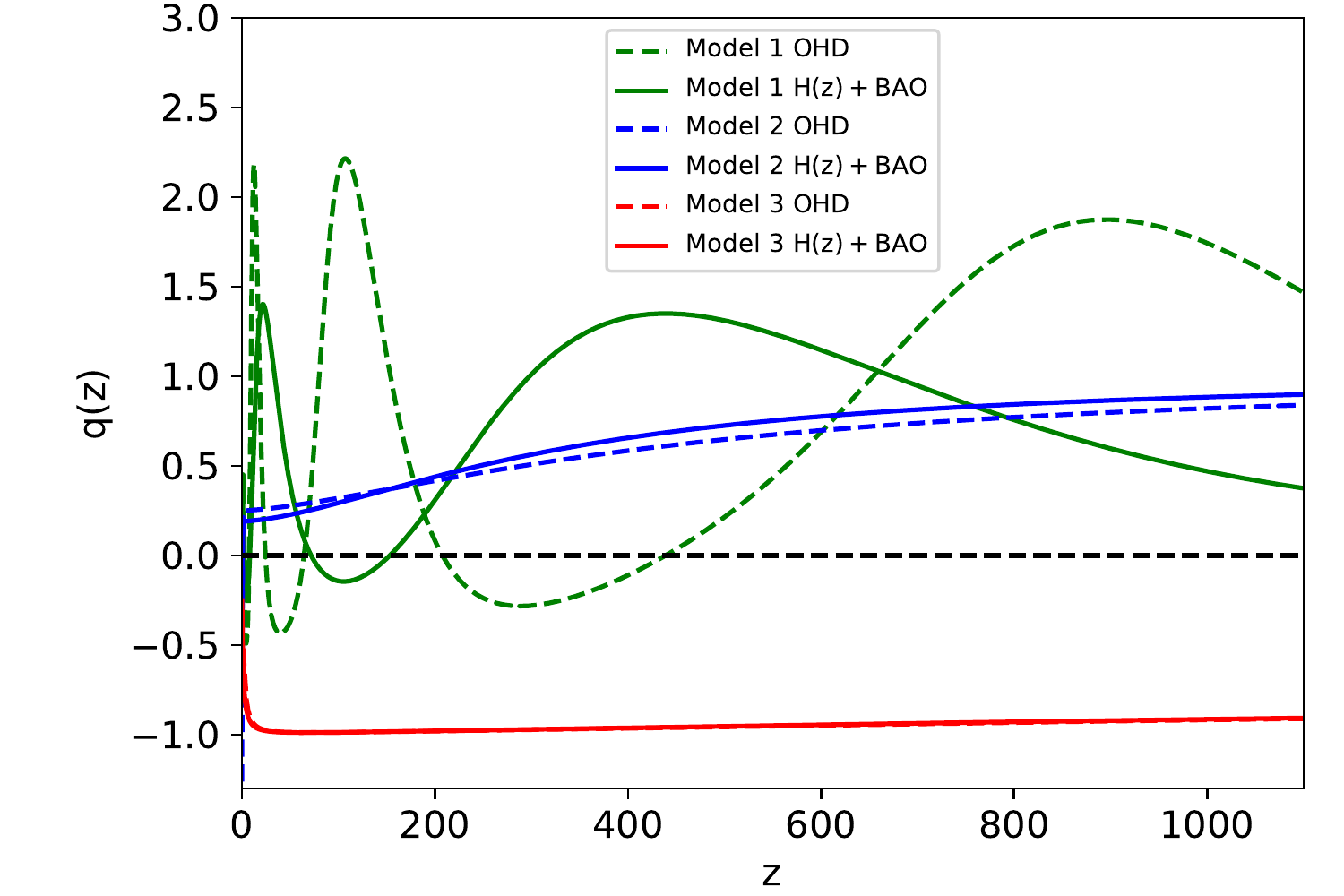}}
  \subfloat[]{%
    \includegraphics[width=3.5in,height=3.0in]{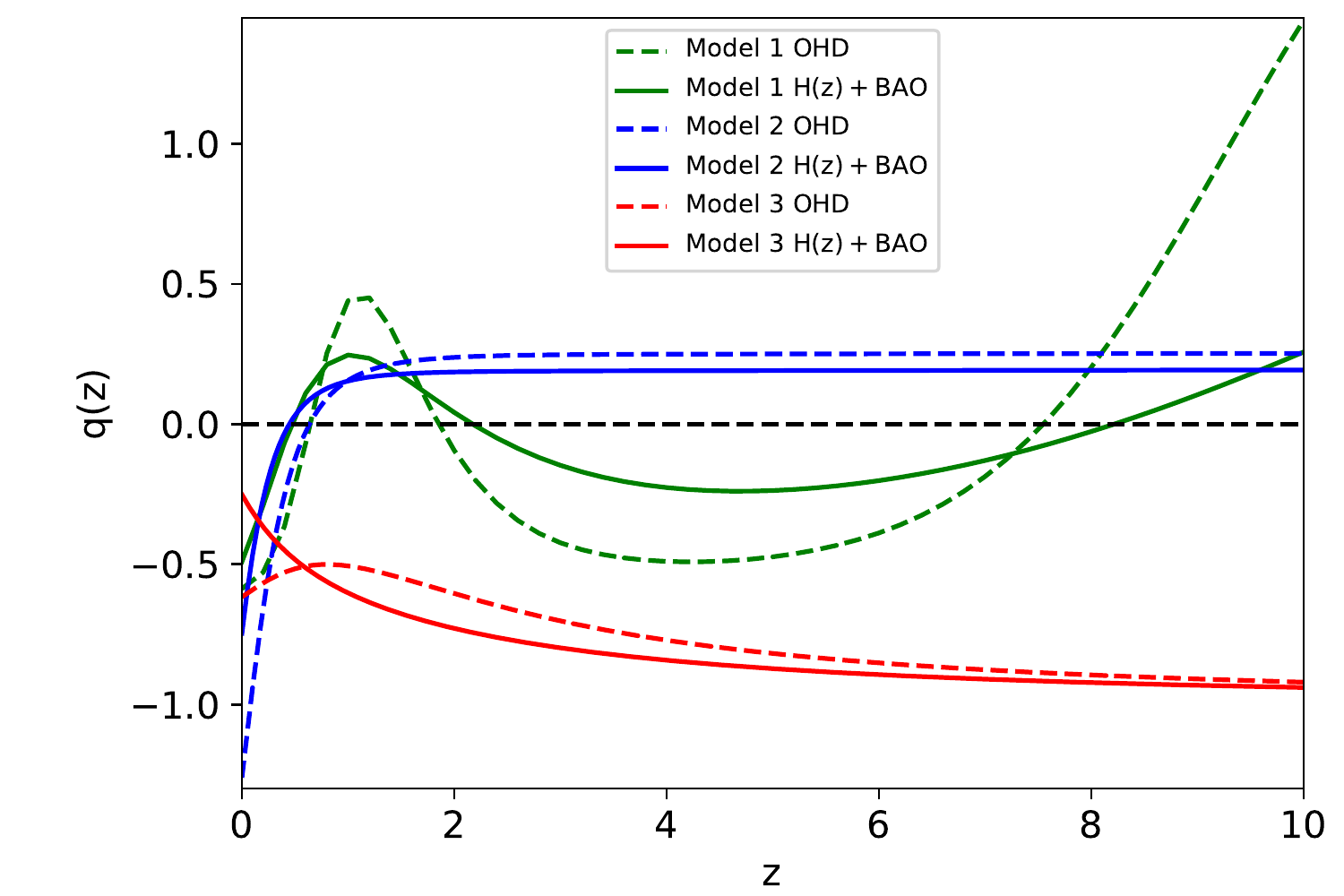}}
  \caption{The deceleration parameter $q(z)$ as a function of redshift $z$ and the right panel shows that of redshift range $0\leq z\leq 10$.}
\label{fig8}
\end{figure}

\end{paracol}
\begin{paracol}{2}
\switchcolumn

\newpage
\end{paracol}
\begin{paracol}{2}
\switchcolumn

\newpage
\end{paracol}
\begin{paracol}{2}
\switchcolumn

\newpage
\end{paracol}
\begin{paracol}{2}
\switchcolumn
\vspace{-6pt}

 \section{Preferred Best-Fitting Models and Their~Implications}
 \label{sec4}
From Table~\ref{table2} we can see that although Model 1 has the lowest $\chi^2_{\rm min}$, based on $AIC$ and $BIC$, Model 2 is the best candidate among these three models. The~evidence against Model 1 is not strong, so Model 1 is not ruled out from these three candidate models. The~values of reduced $\chi^2_{\rm min}\sim0.5$ suggest that the error bars of these data are overestimated. Below~we summarize the main results for the best-fitting models with the unmarginalized best-fitting parameters listed in Table~\ref{table2} in different data combinations. Note that unmariginalized best-fitting parameters means that the parameters are given by maximizing the full likelihood, while the marginalized likelihood means the likelihood is obtained by marginalizing over one (or some) of the parameters.
\begin{description}
  \item[Model 1] With the best-fitting parameters in this model, we~have:
                \begin{enumerate}
                  \item EoR ($6\lesssim z\lesssim15$), for~OHD ($H(z)$ + BAO), from~around 379.9 Myr (402.6 Myr) to 1.825 Gyr (1.718 Gyr), which lasts $\sim$ 1.445 Gyr (1.315 Gyr);
                  \item The seed redshift of J0100+2802, for~OHD ($H(z)$ + BAO), $z_{\rm seed}\sim$ 9.74 (11.03);
                  \item If the first star formed at $z\approx 20$, then the age of the universe at that time would be, for~OHD ($H(z)$ + BAO), 272.8 Myr (234.1 Myr);
                  \item Old globular cluster M92 (NGC 6341) would appear at $z\sim7.53\ (9.27)$ with age of 14.0 Gyr for OHD ($H(z)$ + BAO);
                  \item For OHD ($H(z)$ + BAO), 3.5-Gyr-old ratio galaxy 53W091 ($z\sim1.55$) and 4-Gyr-old radio galaxy 53W069 ($z\sim1.43$) are formed at $z\sim4.98\ (5.40)$ and $z\sim5.41\ (5.99)$, respectively;
                  \item {For OHD ($H(z)$ + BAO),} QSO APM 08279+5255 with age around 2.1 Gyr~\cite{Friaca2005} would have formed at $z\sim9.37\ (12.32)$.
                \end{enumerate}
  \item[Model 2] With the best-fitting parameters in this Model, we~have:
                \begin{enumerate}
                  \item EoR ($6\lesssim z\lesssim15$), for~OHD ($H(z)$ + BAO), from~around 478.7 Myr (534.1 Myr) to 1.371 Gyr (1.462 Gyr), which lasts $\sim$ 892.4 Myr (927.8 Myr);
                  \item The seed redshift of J0100+2802, for~OHD ($H(z)$ + BAO), $z_{\rm seed}\sim17.75\ (16.46)$;
                  \item If the first star formed at $z\approx 20$, then the age of the universe at that time would be, for~OHD ($H(z)$ + BAO), 337.1 Myr (381.4 Myr);
                  \item Old globular cluster M92 (NGC 6341) would appear at $z\sim18.19\ (18.98)$ with age of 14.0 Gyr {for OHD ($H(z)$ + BAO)};
                  \item {For OHD ($H(z)$ + BAO),} 3.5-Gyr-old ratio galaxy 53W091 ($z\sim1.55$) and 4-Gyr-old radio galaxy 53W069 ($z\sim1.43$) are formed at $z\sim5.95\ (6.20)$ and $z\sim6.85\ (7.22)$, respectively;
                  \item {For OHD ($H(z)$ + BAO),} QSO APM 08279+5255 with age around 2.1 Gyr would have formed at $z\sim92.16\ (44.83)$. If~we consider it to be formed after $z\approx 20$, then its age would be $\sim1.810\ (1.861)$ Gyr for OHD ($H(z)$ + BAO).
                \end{enumerate}
  \item[Model 3] With the best-fitting parameters in this model, we~have:
                \begin{enumerate}
                  \item EoR ($6\lesssim z\lesssim15$), for~OHD ($H(z)$ + BAO), from~around 286.8 Myr (293.9 Myr) to 992.2 Myr (1.011 Gyr), which lasts $\sim$ 705.4 Myr (716.8 Myr);
                  \item The seed redshift of J0100+2802, for~OHD and $H(z)$ + BAO, $z_{\rm seed}>20$;
                  \item If the first star formed at $z\approx 20$, then the age of the universe at that time would be, for~OHD ($H(z)$ + BAO), 190.5 Myr (195.3 Myr);
                  \item Old globular cluster M92 (NGC 6341) would appear at $z\sim71.70\ (>1100)$ with age of 14.0 Gyr {for OHD and $H(z)$ + BAO};
                  \item added{For OHD ($H(z)$ + BAO),} 3.5-Gyr-old ratio galaxy 53W091 ($z\sim1.55$) and 4-Gyr-old radio galaxy 53W069 ($z\sim1.43$) are formed at $z\sim6.48\ (6.92)$ and $z\sim7.76\ (8.54)$, respectively;
                  \item For OHD and $H(z)$ + BAO, QSO APM 08279+5255 with age around 2.1 Gyr would have formed at $z>1100$. (ruled out)
                \end{enumerate}
\end{description}

\end{paracol}
\nointerlineskip
\begin{specialtable}[H]
\widetable
\begin{threeparttable}
\caption{Unmarginalized best-fitting parameter values for the coupling~models.}\label{table2}
\setlength{\tabcolsep}{0.7mm}{
\begin{tabular}{lcccccccccccccccc}
\toprule
\textbf{Model} & \textbf{Data Set} & \boldmath{$\Omega_{c_0}\!h^2$} & \boldmath{$\Omega_{b_0}\!h^2$} & \boldmath{$\Omega_{m_0}$} & \boldmath{$w_{\mathrm{X}}$} & \boldmath{$H_0$\textbf{\tnote{a}}} & \boldmath{$A_1$} & \boldmath{$A_2$} & \boldmath{$\lambda$} & \boldmath{$\chi^2_{\rm min}$} & \boldmath{$\nu$\textbf{\tnote{b}}} & \boldmath{$\chi^2_{\rm min}/\nu$} & \boldmath{$AIC$} & \boldmath{$BIC$} & \boldmath{$\Delta AIC$} & \boldmath{$\Delta BIC$} \\
\midrule
Model 1 & OHD & -- & -- & 0.020 & $-0.740$ & 71.34 & 1.210 & 0.333 & -- & 15.85 & 35 & 0.45 & 25.85 & 34.30 & 0.35 & 2.04 \\
 & $H(z)$ + BAO & 0.0010 & 0.0742 & 0.171 & $-0.797$ & 66.64 & 0.907 & 0.472 & -- & 17.85 & 36 & 0.50 & 29.85 & 40.27 & 1.43 & 3.16 \\
\midrule
Model 2 & OHD & -- & -- & 0.383 & $-1.904$ & 76.39 & -- & -- & 0.167 & 17.50 & 36 & 0.49 & 25.50 & 32.26 & 0.00 & 0.00 \\
 & $H(z)$ + BAO & 0.2598 & 0.0064 & 0.579 & $-1.973$ & 67.90 & -- & -- & 0.206 & 18.42 & 37 & 0.50 & 28.42 & 37.11 & 0.00 & 0.00 \\
\midrule
Model 3 & OHD & -- & -- & 0.157 & $-1.107$ & 70.48 & -- & -- & 0.112 & 18.47 & 36 & 0.51 & 26.47 & 33.23 & 0.97 & 0.97 \\
 & $H(z)$ + BAO & 0.1330 & 0.0195 & 0.361 & $-0.607$ & 65.12 & -- & -- & $-0.088$ & 19.65 & 37 & 0.53 & 29.65 & 38.34 & 1.23 & 1.23 \\
\bottomrule
\end{tabular}}
\pbox{\columnwidth}{\footnotesize \mbox{\tnote{a} \hunit. \tnote{b} Degrees of freedom.}}
\end{threeparttable}
\end{specialtable}
\begin{paracol}{2}
\switchcolumn


\clearpage
\end{paracol}
\begin{specialtable}[H]
\widetable
\begin{threeparttable}
\caption{One-dimensional marginalized best-fitting parameter values and uncertainties ($\pm 1\sigma$ or $2\sigma$ upper limits) for the coupling~models.}\label{table3}
\setlength{\tabcolsep}{0.32mm}{
\begin{tabular}{lccccccccc}
\toprule
\textbf{Model} & \textbf{Data Set} & \boldmath{$\Omega_{c_0}\!h^2$} & \boldmath{$\Omega_{b_0}\!h^2$} & \boldmath{$\Omega_{m_0}$} & \boldmath{$w_{\mathrm{X}}$} & \boldmath{$H_0$\textbf{\tnote{a}}} & \boldmath{$A_1$} & \boldmath{$A_2$} & \boldmath{$\lambda$} \\
\midrule
Model 1 & OHD & -- & -- & $0.249^{+0.042}_{-0.029}$ & $-1.077^{+0.347}_{-0.223}$ & $71.28^{+3.95}_{-5.85}$ & $0.376^{+0.220}_{-0.243}$ & $4.521^{+1.827}_{-4.295}$ & -- \\
 & $H(z)$ + BAO & $0.0832^{+0.0402}_{-0.0300}$ & $0.0361^{+0.0069}_{-0.0163}$ & $0.278^{+0.045}_{-0.039}$ & $-0.734^{+0.219}_{-0.095}$ & $65.50^{+2.20}_{-2.86}$ & $0.359^{+0.251}_{-0.215}$ & $4.771^{+2.336}_{-4.510}$ & -- \\
\midrule
Model 2 & OHD & -- & -- & $0.385^{+0.093}_{-0.111}$ & $-2.298^{+1.355}_{-0.813}$ & $77.42^{+6.66}_{-11.21}$ & -- & -- & $0.158^{+0.107}_{-0.019}$ \\
 & $H(z)$ + BAO & $0.1899^{+0.1267}_{-0.1371}$ & $<0.0628$ & $0.444^{+0.183}_{-0.240}$ & $-1.650^{+1.203}_{-0.396}$ & $68.04^{+2.62}_{-5.45}$ & -- & -- & $0.127^{+0.124}_{-0.049}$ \\
\midrule
Model 3 & OHD & -- & -- & $0.363^{+0.255}_{-0.111}$ & $-0.754^{+0.471}_{-0.439}$ & $71.36^{+3.93}_{-4.92}$ & -- & -- & $-0.293^{+0.529}_{-0.230}$ \\
 & $H(z)$ + BAO & $<0.2532$ & $0.0314^{+0.0094}_{-0.0264}$ & $0.344^{+0.065}_{-0.168}$ & $-0.629^{+0.220}_{-0.306}$ & $65.80^{+2.38}_{-2.64}$ & -- & -- & $-0.108^{+0.235}_{-0.059}$ \\
\bottomrule
\end{tabular}}
\pbox{\columnwidth}{\footnotesize \mbox{\tnote{a} \hunit.}}
\end{threeparttable}
\end{specialtable}
\begin{paracol}{2}
\switchcolumn


 Therefore, the~Model 1 OHD and $H(z)$ + BAO cases are the best cases to alleviate the aging problems. In~Figure~\ref{fig5}, the~redshifts when $H(z)/H_{\Lambda \rm CDM}(z)\sim 1$ are 0.40, 0.91, and~1.99 (Model 1 OHD), 0.30 and 1.96 (Model 1 $H(z)$ + BAO), 0.40 (Model 2 OHD), 0.04, 0.24, and~1.83 (Model 2 $H(z)$ + BAO), 0.52 and 725.79 (Model 3 OHD), and~0.16, 1.56, and~984.77 (Model 3 $H(z)$ + BAO); and the current values of $H/H_{\Lambda \rm CDM}$ are 0.49 (Model 1 OHD), 0.33 (Model 1 $H(z)$ + BAO), 0.49 (Model 2 OHD), 0.47 (Model 2 $H(z)$ + BAO), 1.03 (Model 3 OHD), and~1.01 (Model 3 $H(z)$ + BAO). Note that the evolutions of dust matter density parameter $\Omega_m$ and DE density parameter $\Omega_{\rm X}$ are shown in \mbox{Figures~\ref{fig6} and \ref{fig7}} up to $z=1100$, where for Model 1 OHD ($H(z)$ + BAO), the~redshifts when $\Omega_{m}$ have local extrema are 1.11, 4.26, 12.51, 41.89, 107.78, 356.20, and~928.41 (Model 1 OHD); 1.03, 4.72, 21.71, 115.98, and~455.68 (Model 1 $H(z)$ + BAO); 3.98 (Model 2 OHD); 3.23 (Model 1 $H(z)$ + BAO); 14.15 (Model 1 OHD); and 24.84 (Model 1 $H(z)$ + BAO); while the redshifts when $\Omega_{\rm X}$ have local extrema are 1.11, 4.24, 12.51, 40.53, 107.27, 300.45, and~906.09 (Model 1 OHD); 1.03, 4.70, 21.67, 108.17, and~443.89 (Model 1 $H(z)$ + BAO); and none for the rest cases. Since the universe has a slightly larger age in these best-fitting models, we also care about how it works with the expansion of the universe. The~deceleration parameter is
 \beq
 q(z)=-(\frac{\ddot{a}}{a})/H^{2}=\frac{1}{H(z)}\frac{dH(z)}{dz}(1+z)-1.
 \label{qz}
 \eeq
 As shown in Figure~\ref{fig8}, except~for the Model 3 OHD and $H(z)$ + BAO cases, the~deceleration parameter $q(z)$ for the other four cases are going through transitions from deceleration to acceleration phases. For~the Model 1 OHD case, the~transition redshifts (for $z\leq 1100$) are 8.67, 56.70, and~169.15; for the Model 1 $H(z)$ + BAO case, the~transition redshifts are 8.67, 56.70, and~169.15; for the Model 2 OHD case, the~transition redshift is 0.45; and for the Model 2 $H(z)$ + BAO case, the~transition redshift is 0.64; while for the Model 3 OHD and $H(z)$ + BAO cases, $q(z)<0$ is satisfied until~today.

Although the Model 1 OHD and $H(z)$ + BAO cases are capable of explaining the aging issues, Model 1 have some strange (non-physical) behaviors due to the unbounded time-dependent choice of $\lambda$, where the matter density parameter $\Omega_{m}$ and the DE density parameter $\Omega_{\rm X}$ oscillate with the value of $\Omega_{m}$ and $\Omega_{\rm X}$ being a high of 2.51 (1.74) and a low of $-1.52$ ($-0.75$) for the Model 1 OHD and $H(z)$ + BAO cases, respectively. From~panel (b) of Figure~\ref{fig1} we see that the Model 2 cases have relatively larger ages during the evolution and find that other than the age of QSO APM 08279+5255, the~other age issues are significantly improved. Furthermore, judging from $AIC$ and $BIC$ listed in Table~\ref{table2}, Model 2 is the most favored model among these three models. Therefore, we conclude that Model 2 is the best~model.

\newpage
\end{paracol}
\begin{paracol}{2}
\switchcolumn
\vspace{-18pt}

\end{paracol}
\begin{paracol}{2}
\switchcolumn

\newpage
\end{paracol}
\begin{paracol}{2}
\switchcolumn

\end{paracol}
\begin{paracol}{2}
\switchcolumn
\vspace{-6pt}

 \section{Conclusions and~Discussions}
 \label{sec5}
By exploring the properties of the coupling models with the best-fitting parameters obtained from constraints over OHD and $H(z)$ + BAO, we manage to improve the aging problems of the \lcdm\ model, where four of the six best-fitting models demonstrate good improvements on them. Although~the Model 1 OHD and $H(z)$ + BAO cases have better capability of improving the issues, Model 1 has some strange (non-physical) behaviors (see, e.g.,~Figures~\ref{fig6} and \ref{fig7}). The~Model 2 OHD and $H(z)$ + BAO cases can alleviate most of the aging problems and based on $AIC$ and $BIC$ listed in Table~\ref{table2}, Model 2 is considered to be the best candidate model amongst these three models. For~Model 2 OHD ($H(z)$ + BAO), the~constraints on the parameters are $\Omega_{m_0}=0.385^{+0.093}_{-0.111}\ (0.444^{+0.183}_{-0.240})$, $w_{\rm X}=-2.298^{+1.355}_{-0.813}\ (-1.650^{+1.203}_{-0.396})$, $H_0=77.42^{+6.66}_{-11.21}\ (68.04^{+2.62}_{-5.45})$ \hunit, and~$\lambda=0.158^{+0.107}_{-0.019}\ (0.127^{+0.124}_{-0.049})$. The~Model 2 OHD and $H(z)$ + BAO cases support the currently accelerating expansion of the the Universe with the expansions transitioned from decelerating to accelerating at redshits 0.45 and 0.64, respectively. The~ages of the universes are 14.385 Gyr and 14.409 Gyr in the Model 2 OHD and $H(z)$ + BAO cases, respectively. That Model 3 is ruled out does not convey the idea that the coupling between DE and DM is not worth investigating, because~we do not include perturbations in this paper.

\vspace{6pt}



\authorcontributions{Conceptualization, S.C. and T.-J.Z.; writing---original draft preparation, S.C. and X.W.; analysis and plotting, S.C. and T.Z.; writing---review and editing, T.-J.Z., X.W. and T.Z. All authors have read and agreed to the published version of the~manuscript.}

\funding{This work was funded by the National Science Foundation of China (Grants No.11929301, 61802428, 11573006) and National Key R\&D Program of China (2017YFA0402600).}

\institutionalreview{Not applicable.
}

\informedconsent{Not applicable.

}

\dataavailability{Not applicable.}

\acknowledgments{We thank the anonymous referees for insightful suggestions that improved the paper significantly, and~Qiang Xu and Zhongxu Zhai for useful ideas and~discussions.}

\conflictsofinterest{The authors declare no conflict of~interest.}






\end{paracol}
\reftitle{References}

\end{document}